\documentclass[reprint,twocolumn,secnumarabic,amssymb, aps, prl]{revtex4-2}

\newcommand{\naturepar}[1]{\vspace{10pt}\noindent\textbf{#1}\\}

\usepackage{graphicx}
\usepackage{amsmath,amssymb}
\usepackage{braket}
\usepackage[colorlinks,linkcolor=blue,citecolor=blue,urlcolor=blue,filecolor=black]{hyperref}

\usepackage{tikz}
\usepackage[dvipsnames]{xcolor}
\usepackage{amsthm}
\usepackage{enumerate}

\usetikzlibrary{arrows.meta, positioning, calc, fit, backgrounds, decorations.pathreplacing, patterns}

\definecolor{deepslate}{HTML}{355070}
\definecolor{warmrust}{HTML}{B7653A}
\definecolor{sage}{HTML}{81B29A}
\definecolor{wheat}{HTML}{F2CC8F}
\definecolor{ocean}{HTML}{3D7EA6}
\definecolor{mauve}{HTML}{6D597A}

\tikzset{
    pipelinebox/.style={
        draw=deepslate!50, fill=deepslate!5,
        rounded corners=3pt, minimum height=1.0cm, minimum width=2.0cm,
        font=\sffamily\small\bfseries, align=center,
        inner sep=5pt, line width=0.7pt,
    },
    convbox/.style={
        draw=warmrust!60, fill=warmrust!8,
        rounded corners=3pt, minimum height=1.0cm, minimum width=2.0cm,
        font=\sffamily\small\bfseries, align=center,
        inner sep=5pt, line width=0.7pt,
    },
    pipearrow/.style={
        -{Stealth[length=6pt, width=4.5pt]},
        line width=0.9pt, deepslate!60,
    },
    branchline/.style={
        line width=0.7pt, deepslate!35, densely dashed,
    },
    geolabel/.style={
        font=\sffamily\small\bfseries, align=center, text=deepslate!80,
    },
    geosublabel/.style={
        font=\sffamily\scriptsize, align=center, text=deepslate!50,
    },
    panellabel/.style={
        font=\sffamily\bfseries\normalsize,
    },
    paneltitle/.style={
        font=\sffamily\small, text=deepslate!70,
    },
}

\usepackage{setspace}
\usepackage[export]{adjustbox}
\usepackage{dsfont}
\usepackage[capitalize]{cleveref}
\crefname{appendix}{Appendix}{Appendices}
\Crefname{appendix}{Appendix}{Appendices}
\usepackage{bm}
\usepackage{mathtools}
\usepackage{url}
\usepackage{enumitem}
\usepackage{verbatim}
\usepackage{MnSymbol}
\usepackage{physics}
\usepackage{stmaryrd}
\usepackage{lineno}

\begin{document}

\title{Scalable Neural Decoders for Practical Fault-Tolerant Quantum Computation}

\author{Andi~Gu,$^{1,2,*}$ J.~Pablo~Bonilla~Ataides,$^{1,2}$ Mikhail~D.~Lukin,$^{1,2}$ and Susanne~F.~Yelin$^{1,2}$}

\affiliation{\mbox{$^1$Department~of~Physics,~Harvard~University,~Cambridge,~MA~02138,~USA}\\\mbox{$^2$Harvard~Quantum~Initiative,~Harvard~University,~Cambridge,~MA~02138,~USA}\\$^*$andigu@g.harvard.edu}

\begin{abstract}
Quantum error correction (QEC) is essential for scalable quantum computing. 
However, it requires 
classical decoders that are fast and accurate enough to keep pace with quantum hardware. While quantum low-density parity-check codes have recently emerged  as a promising route to efficient fault tolerance~\cite{leverrier2022quantum,panteleev2022quantum,bravyi2024high}, current decoding algorithms do not allow one to realize the full potential of these codes in practical settings~\cite{poulin2008iterative,roffe2020decoding,shutty2025tesseract}.
Here, we introduce 
a convolutional neural network decoder that exploits the geometric structure of QEC codes, and use it to probe a novel ``waterfall" regime of error suppression, demonstrating  
that the logical error rates required for large-scale fault-tolerant algorithms~\cite{gidney2021factor,beverland2022assessing} are attainable with modest code sizes at current physical error rates~\cite{ballance2016high,li2024realization,evered2025highfidelity}, and with latencies within the real-time budgets of several leading hardware platforms. For example, 
for the $\llbracket 144, 12, 12 \rrbracket$ Gross code~\cite{bravyi2024high}, the decoder achieves logical error rates up to ${\sim}17\times$ below existing decoders~\cite{roffe2020decoding,muller2024relax,shutty2025tesseract}---reaching logical error rates $\sim 10^{-10}$ at physical error $p = 0.1\%$---with 3-5 orders of magnitude  
higher throughput.
This decoder also produces well-calibrated confidence estimates that can significantly reduce the time overhead of repeat-until-success protocols~\cite{campbell2017roads,smith2024mitigating}.
Taken together, these results suggest that the space-time costs associated with fault-tolerant quantum computation may be significantly lower than previously anticipated.
\end{abstract}
\maketitle

\begin{figure*}
    \centering
    \includegraphics[width=2\columnwidth]{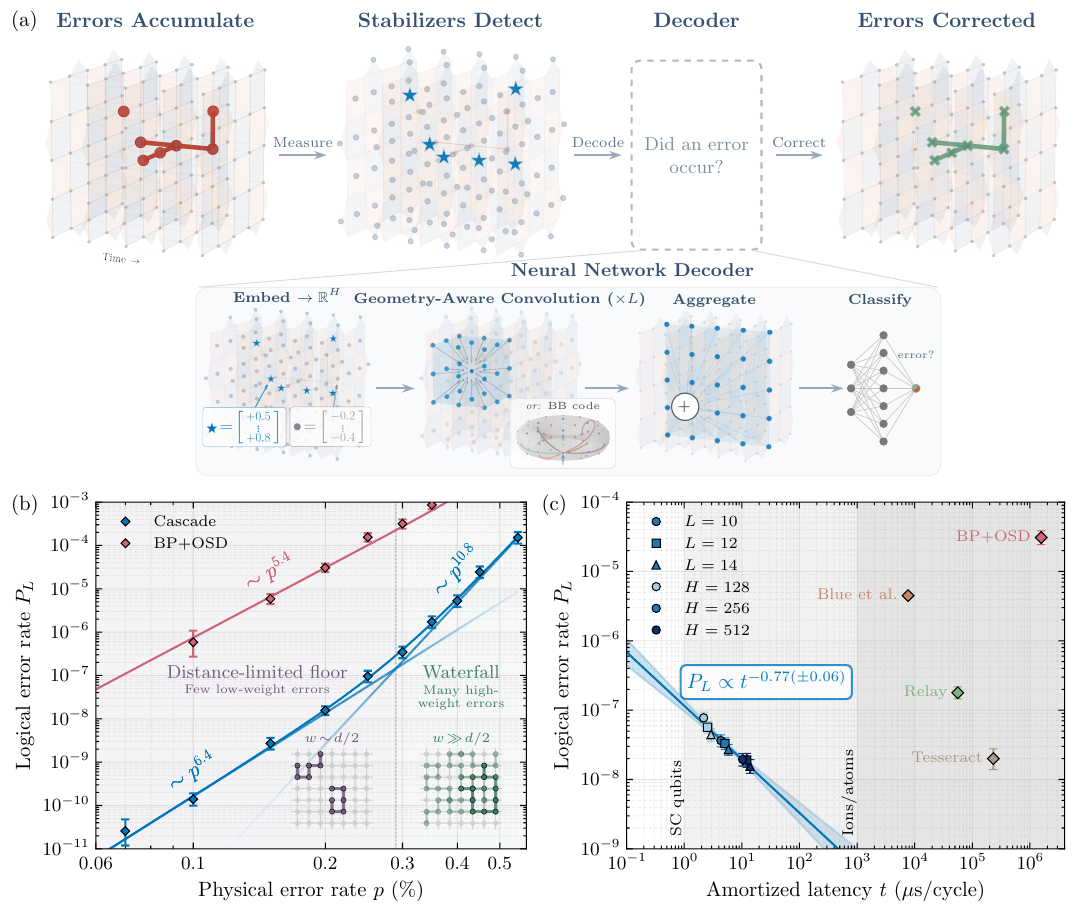}
    \caption{\textbf{Structure-aware neural decoding for quantum error correction.}
    (a)~Top: errors accumulate on data qubits; stabilizer measurements produce a spacetime syndrome (pattern of detection events); a decoder determines whether a logical error occurred; the logical state is protected.
    Bottom (neural network decoder): syndromes are embedded into $H$-dimensional representations and processed by $L$ convolutional layers whose local structure respects the code geometry---3D convolutions for surface codes, generalized convolutions on the torus for bivariate bicycle (BB) codes---followed by a final convolution scattering to data qubits, pooling over the data qubits in each logical operator's support, and a prediction head applied independently to each logical observable.
    (b)~Error suppression on the $\llbracket 144, 12, 12 \rrbracket$ BB code ($R = d$ rounds). The logical error rate per logical qubit per cycle, $P_L \approx \sum_w N(w)\,p^w$, where $N(w)$ is the number of minimal failure modes of weight $w$, decomposes into two regimes: a steep waterfall ($\sim p^{10.8}$) where the numerous high-weight failure modes dominate at moderate physical error rates $p$, transitioning to a distance-limited floor ($\sim p^{6.4}$) at very low noise. BP+OSD (orange, $\sim p^{5.4}$) misses the waterfall entirely.
    (c)~Accuracy--latency tradeoff at $p=0.2\%$: Cascade (GPU inference on NVIDIA H200) spans a range of (amortized) latencies while achieving lower logical error rates than prior decoders (single-threaded CPU). Diamond markers show the reported single-shot latencies of BP+OSD~\cite{roffe2020decoding}, Relay~\cite{muller2024relax}, Tesseract~\cite{shutty2025tesseract}, and the ML decoder of~\cite{blue2025machinelearningdecodingcircuitlevel}. Error bars indicate 95\% credible intervals.}
    \label{fig:hero}
\end{figure*}

Quantum computers promise to solve problems beyond the reach of classical machines~\cite{feynman1982simulating,shor1997polynomial,aspuru2005simulated,gidney2021factor}. However, this is possible only if the inherent fragility of large scale quantum systems can be circumvented  through the use of error correction.
Quantum error correction (QEC) codes suppress errors exponentially as the code size increases---provided the physical error rate lies below a critical threshold~\cite{aharonov1997fault,knill1998resilient}.
Recent experiments have demonstrated this error suppression on superconducting hardware~\cite{google2021exponential,google2024willow}, neutral atom platforms~\cite{bluvstein2024logical,bluvstein2026fault}, and trapped-ion systems~\cite{paetznick2024logical}, marking significant progress toward fault-tolerant quantum computation.
However, achieving the logical error rates required for realization of utility-scale 
quantum 
algorithms~\cite{gidney2021factor,campbell2017roads} remains a major challenge. 

While recent experiments have used topological QEC codes such as surface and color codes~\cite{fowler2012surface,bombin2006topological},
a new generation of quantum LDPC codes have recently emerged that feature a combination of  high encoding rates and large distances~\cite{leverrier2022quantum,panteleev2022quantum,panteleev2022asymptotically,bravyi2024high} and offer a path toward more efficient fault tolerance.
Deploying these codes in practice requires a classical decoder that is simultaneously fast and accurate, interpreting syndrome measurements and identifying corrections in real time.
For quantum LDPC codes, no existing decoder currently meets both requirements.
Belief propagation, the standard iterative algorithm, suffers from convergence failures due to trapping sets~\cite{poulin2008iterative,raveendran2023trapping} (patterns that cause it to converge to incorrect solutions), leaving logical error rates orders of magnitude worse than what can be achieved theoretically; on the other hand, the most accurate methods are far too slow for real-time use~\cite{panteleev2021degenerate,roffe2020decoding,shutty2025tesseract,gu2026colorcodesurfacecode}.
A similar gap emerged in classical error correction, where Gallager's LDPC codes achieved near-Shannon-limit performance in theory but remained largely unused for three decades until practical iterative decoders transformed them into the foundation of modern communications~\cite{gallager1962low,mackay1997near}.
Moreover, the absence of an efficient decoder may also obscure the potential power of these codes. While 
the  resource estimates for fault-tolerant quantum computation  widely model logical error rates using QEC threshold and the code distance $d$, using 
the assumption that minimum-weight uncorrectable errors dominate~\cite{gidney2021factor,litinski2019game,beverland2022assessing,campbell2017roads,zhou2025resource}, in classical error correction  high-rate codes can sometimes   
suppress errors more effectively than suggested by this worst-case bound~\cite{richardson2001capacity,richardson2009modern}.

Here we introduce Cascade, a convolutional neural network decoder, that exploits the geometric regularity of QEC codes. Unlike belief propagation, whose fixed update rules cause it to converge to incorrect solutions on degenerate error patterns, the neural decoder learns flexible message-passing rules that can circumvent these failure modes. Using this decoder, we show that, rather remarkably,  error suppression on both bivariate bicycle (BB) codes~\cite{bravyi2024high} and surface codes can significantly exceed the naive distance scaling.  This is because the code distance $d$ identifies the lowest weight uncorrectable error, but says nothing about how \emph{likely} such an error is; minimum-weight uncorrectable patterns are typically very rare, so that below threshold, higher-weight failure modes, individually less probable but vastly more numerous, dominate the failure rate. The resulting two-regime structure, with a steep ``waterfall'' in error rate below threshold that reverts to distance-limited scaling at very low noise, is well established for classical LDPC codes~\cite{richardson2001capacity,richardson2009modern}. Indications of similar behavior have appeared in quantum codes~\cite{watson2014logical,kasai2012quantum,komoto2025qec}, but existing decoders lack the accuracy to resolve it in the practically relevant regime below threshold.
Our results demonstrate that quantum LDPC codes exhibit a dramatic waterfall scaling. For instance, we find that for the $\llbracket 144, 12, 12 \rrbracket$ Gross code, the logical error rate is well described by two power-law contributions: a steep waterfall term $P_L\sim p^{11}$ and a distance-limited term $P_L\sim p^{6.4} \approx p^{\lfloor (d+1)/2 \rfloor}$ (\cref{fig:hero}(b)).
The large gap between these exponents reflects properties of the code's error structure beyond the minimum distance, suggesting that distance alone is an incomplete predictor of practical performance.
Moreover, Cascade achieves near-optimal accuracy at a fraction of the computational cost, surpassing all practical decoders by orders of magnitude on the Gross code at $p=0.1\%$, while achieving near-optimal performance on surface codes. This decoding method is also very practical. Its local, feed-forward architecture is well suited to hardware acceleration, with GPU latencies already within the budgets for trapped-ion and neutral-atom platforms~\cite{bluvstein2024logical}.
Models trained at a single high noise level generalize reliably across seven orders of magnitude in logical error rate, with no error floor---exponential error suppression persists to $P_L \approx 10^{-11}$.
This generalization extends to the decoder's confidence estimates, which remain well-calibrated across noise levels.
Together, these results reduce both the space and time costs of fault-tolerant quantum computation: the steep waterfall implies that the required code size grows substantially more slowly with target logical error rate than the standard formula predicts, while calibrated confidence estimates reduce the time overhead of repeat-until-success protocols~\cite{smith2024mitigating,zhou2025error,menon2025magic} such as magic state distillation~\cite{campbell2017roads}.

\subsection*{Structure-aware neural decoding} 

In a QEC memory experiment, errors accumulate on data qubits during repeated rounds of stabilizer measurement, producing a spacetime syndrome---the pattern of detection events that the decoder must interpret to determine whether a logical error has occurred (\cref{fig:hero}(a)).
The decoding problem is to infer, from this syndrome, the equivalence class of the error: many distinct physical error patterns produce the same syndrome, so the decoder need only determine whether the net effect is a logical error, not reconstruct the exact error.

The QEC codes we consider---surface codes and bivariate bicycle (BB) codes---possess geometric structure that a decoder can exploit.
Both code families arrange stabilizers in a regular spatial lattice (a 2D grid for surface codes; a torus $\mathbb{Z}_\ell \times \mathbb{Z}_m$ for BB codes), which under repeated rounds of syndrome measurement extends into a spatiotemporal structure where every stabilizer has the same local connectivity up to translation.
This regularity motivates the three design principles of Cascade.
First, \emph{locality}: errors produce spatially localized syndrome patterns that can be progressively resolved by layers of local processing.
Second, \emph{translation equivariance}: the identical local structure at every site means the same decoding rules should apply everywhere.
Third, \emph{anisotropy}: information arriving from different directions carries distinct meaning---for instance, a surface code stabilizer's horizontal and diagonal neighbors are of different type (X versus Z), and the temporal direction encodes measurement errors rather than data qubit errors.
While we focus on surface codes and BB codes, these three properties hold for any stabilizer code whose checks share a common local structure up to spatial symmetries, encompassing a broad class of codes including color codes~\cite{bombin2006topological}, toric codes, hyperbolic surface codes~\cite{breuckmann2021quantum}, and other quantum LDPC codes defined on periodic lattices or Cayley graphs.
For codes with less regularity, stabilizers can be grouped into equivalence classes that share weights, reducing to a standard graph neural network in the fully irregular limit.

Belief propagation satisfies the first of these properties but not the others: it performs local message passing with fixed update rules that treat the graph structure symmetrically, causing it to converge to incorrect solutions on degenerate quantum error patterns~\cite{poulin2008iterative,raveendran2023trapping}.
Our decoder replaces these fixed rules with learned convolutional operations that satisfy all three properties simultaneously (\cref{fig:hero}(a)).
Binary detection events are first embedded into $H$-dimensional representations at each syndrome location.
These representations are then processed by $L$ convolutional layers, where each layer performs learned, direction-dependent message passing between neighboring sites---for surface codes, a standard 3D convolution over the spacetime lattice; for BB codes, a generalized convolution on the torus with weights indexed by the relative offset between stabilizers.
Because the convolutional weights are shared across all positions, the decoder applies identical learned rules at every site, exploiting the translation symmetry of the code while using direction-specific weights to distinguish information arriving from different neighbors.
Successive layers expand the receptive field, so that after $L \sim d$ layers the network integrates information across the full code distance---resolving first local error patterns, then progressively larger-scale correlations, and finally the global topological ambiguity that determines the logical error class. After the final layer, the representations are scattered to data qubits, pooled over the support of each logical operator, and passed to a prediction head that outputs a probability for each logical observable.
The model is trained end-to-end with binary cross-entropy loss at a high physical error rate, requiring no auxiliary losses, multi-stage fine-tuning, or labeled data at low noise levels.
Full architectural and training details are given in Methods.

\subsection*{Error suppression, speed and accuracy}

\begin{figure*}[t]
    \centering
    \includegraphics[width=2\columnwidth]{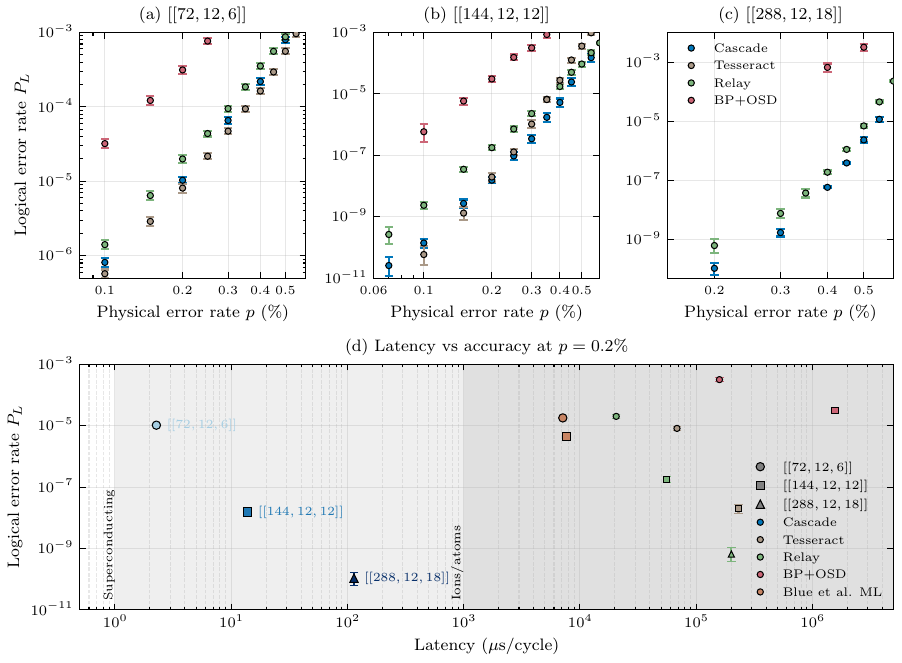}
    \caption{\textbf{Distance scaling of BB code decoders under circuit-level depolarizing noise.}
    (a--c)~Logical error rate per logical qubit per round ($R = d$ rounds) versus physical error rate for the $\llbracket 72, 12, 6 \rrbracket$, $\llbracket 144, 12, 12 \rrbracket$, and $\llbracket 288, 12, 18 \rrbracket$ bivariate bicycle codes, respectively. Cascade achieves lower logical error rates than BP+OSD, Relay, and (where evaluable) Tesseract across all code sizes.
    (d)~Accuracy vs. latency at $p = 0.2\%$ for all three BB codes (timing reported for BB codes is amortized latency). We include the published results of a different ML decoder~\cite{blue2025machinelearningdecodingcircuitlevel} for the $\llbracket 72, 12, 6 \rrbracket$ and $\llbracket 144, 12, 12 \rrbracket$ codes. For the $\llbracket 288, 12, 18 \rrbracket$ code, we are unable to evaluate Tesseract due to its computational cost; BP+OSD is similarly intractable at lower noise levels.}
    \label{fig:bb_distance}
\end{figure*}
\begin{figure*}
    \centering
    \includegraphics[width=2\columnwidth]{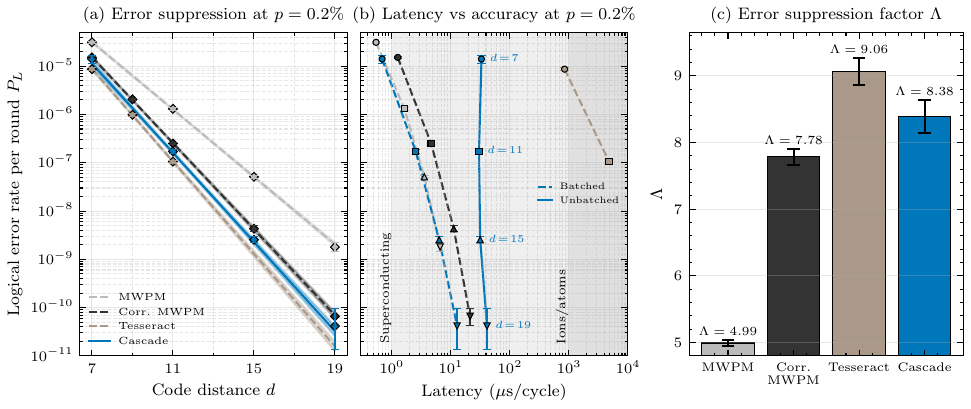}
    \caption{\textbf{Distance scaling of surface code decoders at $p = 0.2\%$ under circuit-level depolarizing noise.}
    (a)~Logical error rate per round versus code distance $d$ for MWPM, correlated MWPM, Tesseract, and Cascade. All decoders exhibit exponential error suppression ($P_L \propto \Lambda^{-\lfloor(d+1)/2\rfloor}$), with Cascade and Tesseract achieving the steepest slopes.
    (b)~Accuracy--latency trade-off across code distances (GPU inference on NVIDIA H200). Amortized latency (dashed lines, batched inference) is significantly lower than the single-shot latency (solid lines, unbatched inference).
    (c)~Error suppression factor $\Lambda$ extracted from exponential fits to panel~(a).}
    \label{fig:distance}
\end{figure*}
\Cref{fig:hero}(b) shows the logical error rate per logical qubit per cycle for the $\llbracket 144, 12, 12 \rrbracket$ Gross code under circuit-level depolarizing noise.
Our largest model ($L=14$, $H=512$) achieves logical error rates orders of magnitude below belief propagation with ordered statistics decoding (BP+OSD)~\cite{roffe2020decoding} and Relay~\cite{muller2024relax} across the full noise range~\footnote{For all reference decoders, we use BP+OSD with default \texttt{stimbposd} settings (product-sum BP, 30 iterations, combination-sweep OSD of order 60)~\cite{roffe2020decoding}, Relay with hyperparameters tuned following the authors' recommended procedure~\cite{muller2024relax}, and Tesseract in the short-beam configuration~\cite{shutty2025tesseract}.}.
Below threshold, the logical error rate drops far more steeply than the naive distance scaling $P_L \sim p^{\lfloor(d+1)/2\rfloor}$ would predict.
For this code ($d=12$), the logical error rate decomposes into two power-law contributions (\cref{fig:hero}(b)): a steep `waterfall' term $\sim p^{11}$ that dominates at moderate error rates, and a distance-limited floor $\sim p^{6.4} \approx p^{\lfloor(d+1)/2\rfloor}$ that emerges at very low noise.
The origin of this decomposition lies in the structure of the code's failure modes.
A minimal failure mode of weight $w$ is a set of $w$ errors that causes a logical failure, with no failing proper subset; to leading order, the logical error rate decomposes as $P_L \approx \sum_{w} N(w)\, p^w$, where $N(w)$ counts the number of such modes.
The standard scaling formula $P_L \sim p^{\lfloor(d+1)/2\rfloor}$ implicitly assumes that the minimum-weight failure modes dominate---and at sufficiently low noise they do, because $p^w$ decreases so steeply with $w$ that the lowest-weight term always wins eventually.
However, for the codes studied here, $N(w)$ is heavily concentrated at weights well above $\lfloor d/2 + 1 \rfloor$: the code admits very few minimal failure modes near the minimum weight, but many more at higher weight.
At moderate physical error rates below threshold, the product $N(w)\,p^w$ is therefore dominated by these higher-weight modes, whose large multiplicity more than compensates for the additional suppression from $p^w$.
The result is two distinct regimes: a steep waterfall where the abundant high-weight failure modes dominate the logical error rate, and a distance-limited floor that only emerges once the physical error rate is low enough for the rare minimum-weight modes to take over.
This two-regime structure is well characterized for classical LDPC codes~\cite{richardson2001capacity,richardson2009modern}, but had not previously been studied  in quantum codes, because the existing quantum decoders are not accurate enough to probe it.
Crucially, the set of minimal failure modes depends on the decoder: a more powerful decoder eliminates failure modes at low weight, exposing the regime where the numerous high-weight modes dominate; decoders like BP+OSD fail on too many low-weight errors, so the steep suppression is never exposed.
As a consequence, our decoder achieves logical error rates ${\sim}4000\times$ below BP+OSD~\cite{roffe2020decoding} and ${\sim}17\times$ below Relay~\cite{muller2024relax} on the Gross code at $p=0.1\%$, with accuracy comparable to Tesseract~\cite{shutty2025tesseract}.
We observe no error floor: exponential error suppression persists to the lowest physical error rates tested ($P_L \approx 2 \times 10^{-11}$).
Unlike matching decoders, which provably correct all errors below weight $\lfloor(d-1)/2\rfloor$, neural decoders offer no such guarantee---systematic misclassification of specific low-weight error patterns could create an irreducible floor analogous to trapping sets in classical BP decoding.
The absence of any floor down to $P_L \approx 2 \times 10^{-11}$ provides strong evidence that the learned decoding rules avoid such failure modes. \Cref{fig:bb_distance}(a--c) extends this analysis to three BB codes---$\llbracket 72, 12, 6 \rrbracket$, $\llbracket 144, 12, 12 \rrbracket$, and $\llbracket 288, 12, 18 \rrbracket$---showing that error suppression improves with code distance from $d=6$ to $d=18$, with our decoder consistently outperforming reference decoders across all three code sizes.
On the largest code ($\llbracket 288, 12, 18 \rrbracket$), the decoder achieves $P_L \sim 10^{-10}$ per logical qubit per cycle at a physical error rate $p = 0.2\%$.

\begin{figure*}
    \centering
    \includegraphics[width=0.9\textwidth]{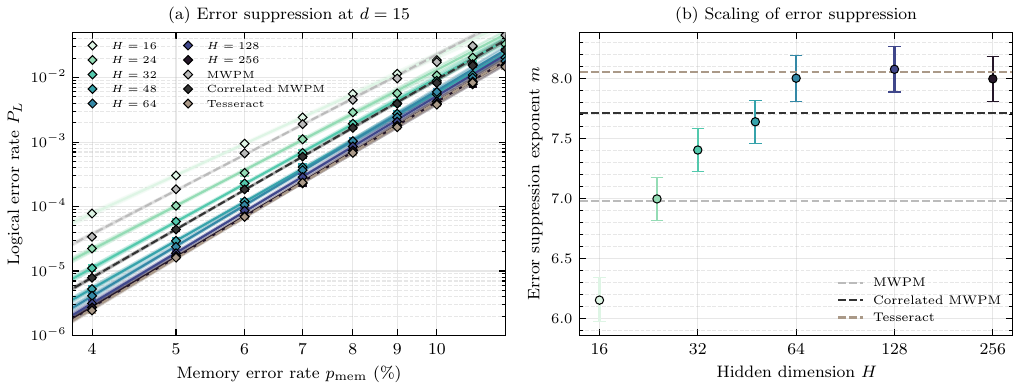}
    \caption{\textbf{Error suppression improves with model capacity.}
    (a) Logical error rate versus memory error rate $p_\mathrm{mem}$ at distance $d=15$ for surface code models with varying hidden dimension $H$ (data-level depolarizing noise, used here for training efficiency). Larger models achieve lower error rates across the full range of noise levels.
    (b) Error suppression exponent $m$ for memory errors (from fitting $P_L \propto p^m$) versus hidden dimension. Small models ($H \lesssim 64$) exhibit poor scaling exponents well below reference decoders (MWPM, correlated matching), while large models ($H \geq 64$) approach optimal performance.}
    \label{fig:emergence}
\end{figure*}

The waterfall is not specific to quantum LDPC codes.
On surface codes, the standard model $P_L \sim (p/p_{\mathrm{th}})^{\lfloor(d+1)/2\rfloor}$ predicts that error suppression is characterized by a single error suppression factor $\Lambda = p_{\mathrm{th}}/p$, determined entirely by the decoder's threshold.
Fitting $P_L \propto \Lambda^{-\lfloor(d+1)/2\rfloor}$ across distances $d = 7$ to $19$ at $p = 0.2\%$ (\cref{fig:distance}), our decoder achieves the nominal value $\Lambda \approx 8.4$, substantially exceeding both uncorrelated minimum-weight perfect matching (MWPM, $\Lambda \approx 5.0$) and correlated MWPM ($\Lambda \approx 7.8$), and approaching Tesseract ($\Lambda \approx 9.1$), a near-optimal decoder whose computational cost (up to $1\,\mathrm{s}$ per shot) makes it impractical for real-time use but serves as a useful performance benchmark.
The nearly twofold variation---from $5.0$ for MWPM to $9.1$ for Tesseract---exceeds what threshold differences alone can account for.
The excess suppression is also a signature of the waterfall: at moderate code distances, the effective error suppression is steeper than $\lfloor(d+1)/2\rfloor$ because minimum-weight uncorrectable errors are rare, and it is the decoder's ability to correctly handle the more numerous higher-weight errors that determines how much of this steeper suppression is realized.
We note that the fit $P_L \propto \Lambda^{-\lfloor(d+1)/2\rfloor}$ remains well-defined because the effective slope is well-approximated by a constant multiple of $\lfloor(d+1)/2\rfloor$ across the distances tested; this multiplicative factor is absorbed into the extracted $\Lambda$, which therefore captures the decoder's total error suppression per unit distance.

These observations have direct implications for resource estimation.
The projections of Refs.~\cite{gidney2021factor,litinski2019game,beverland2022assessing} assume $\Lambda \approx 10$ at $p = 0.1\%$, a value calibrated to MWPM-class decoders; a decoder that achieves $\Lambda$ significantly above this value at the same physical error rate would reduce the code distance---and therefore the physical qubit count---needed to reach any given target logical error rate.
Concretely, our decoder reaches a target logical error rate of $\sim\!10^{-9}$ at code distance $d=15$, compared to $d=19$ for MWPM---a ${\sim}40\%$ reduction in physical qubit count on the code family that underpins current hardware experiments~\cite{google2024willow,bluvstein2024logical}.
Moreover, this advantage grows with stricter targets: because the waterfall gives steeper-than-distance scaling, the gap between waterfall-aware and conventional resource estimates widens as the target logical error rate decreases toward the $10^{-10}$--$10^{-12}$ regime required for large-scale algorithms~\cite{gidney2021factor,campbell2017roads}.

These accuracy gains are only practically relevant if the decoder is fast enough for real-time operation.
\Cref{fig:hero}(c) and \Cref{fig:bb_distance}(d) show the accuracy--latency tradeoff at $p = 0.2\%$ for BB codes, while \Cref{fig:distance}(b) shows how latency scales with code distance for surface codes.
Our models have single-shot latencies of $\sim 40~\mu\mathrm{s}$ per cycle on a single NVIDIA H200 GPU; batched inference achieves amortized latencies up to two orders of magnitude lower at comparable accuracy, for an overall throughput $3{,}000$--$100{,}000\times$ higher than existing decoders (single-threaded CPU)~\footnote{Reported amortized latencies are computed as total wall-clock time divided by (batch size $\times$ syndrome rounds). This includes per-invocation overhead amortized across rounds.}.
Which metric matters depends on the computational setting: single-shot latency is the constraint when mid-circuit measurements gate subsequent operations, while amortized latency is the relevant figure of merit when syndrome rounds can be buffered and decoded in parallel~\cite{terhal2015quantum,skoric2023parallel}, as in memory experiments or deep Clifford circuits.
These latencies are well within the decoding budgets for trapped-ion and neutral-atom platforms ($\sim 1~\mathrm{ms}$)~\cite{bluvstein2024logical}, but above the $\sim 1~\mu\mathrm{s}$ required for superconducting qubits. We note that 
the architecture's local, regular computational structure is well-suited to hardware acceleration and  roofline estimates suggest that FLOP-efficient variants on dedicated hardware could approach the superconducting budget (Supplementary Information).

\subsection*{Robustness and generalization}
\label{sec:emergence}

How well a decoder exploits a code's error-correcting capability depends strongly on its capacity.
To investigate this dependence, we trained surface code decoders of varying width $H$ at a single noise level under data-level depolarizing noise and evaluated across a range of physical error rates (\cref{fig:emergence}).
Small models ($H \lesssim 64$) achieve error suppression worse than uncorrelated MWPM.
As capacity increases, the suppression exponent $m$ (from $P_L \propto p^m$) rises continuously, saturating at $m \approx 8$ for $H \gtrsim 64$---near the optimal value.
Under data-level noise, this optimum coincides with the distance bound; under circuit-level noise, the same capacity dependence applies, but optimal performance includes the steep waterfall regime observed above.
This distinction reflects the combinatorial structure of failure modes: under data-level noise, minimum-weight logical failures correspond to the many shortest paths across the lattice, so these modes dominate and the distance bound is tight; under circuit-level noise, a minimum-weight failure requires a specific alignment of faults in spacetime, making such modes rare relative to the more numerous higher-weight failures that drive the waterfall.
These considerations indicate that insufficient capacity is thus a key reason existing decoders fail to access the waterfall regime: as the model's expressive capacity grows, it gains the ability to recognize and correctly classify increasingly complex error patterns---patterns that simpler decoders systematically mishandle.

\begin{figure*}
    \centering
    \includegraphics[width=2\columnwidth]{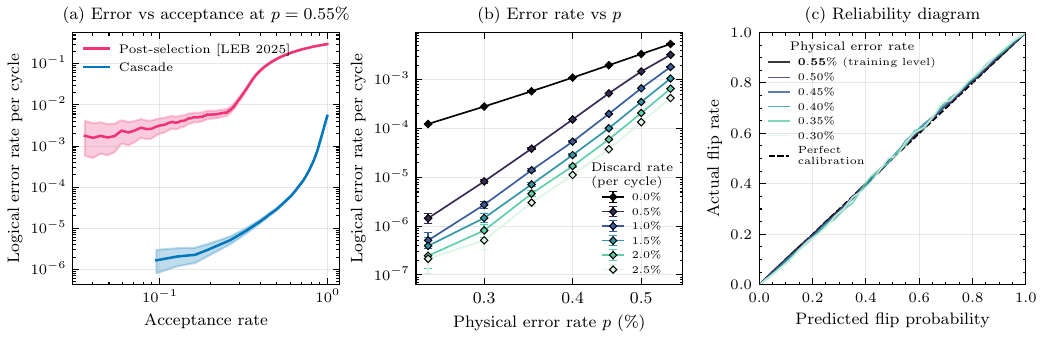}
    \caption{\textbf{Confidence-aware decoding with the ML decoder on the $\llbracket 72, 12, 6 \rrbracket$ bivariate bicycle code.} (a) Logical error rate per cycle as a function of acceptance rate at $p = 0.55\%$. Our decoder (blue) achieves much steeper error suppression (as a function of acceptance rate) compared to cluster-based post-selection methods~\cite{lee2025efficient} (pink). (b) Logical error rate versus physical error rate for different discard rates. (c) Reliability diagram showing calibration across physical error rates (color scale); alignment with the diagonal indicates well-calibrated predictions.}
    \label{fig:calibration}
\end{figure*}

All models are trained at a single high physical error rate, yet generalize reliably across seven orders of magnitude in logical error rate---a generalization that extends beyond accuracy to the network's uncertainty estimates.
\label{sec:calibration}
Although trained at a single noise level, the decoder's predicted probabilities remain well-calibrated across physical error rates far below the training distribution (\cref{fig:calibration}(c)).
This is not guaranteed a priori: the posterior probability of a logical error given a syndrome depends on the physical error rate, which varies across the evaluation range.
A possible explanation is that the syndrome's local structure---the density and clustering of detection events---provides an implicit signal for error severity that scales consistently with the physical error rate, allowing the network's confidence estimates to remain meaningful even far from the training distribution.
Calibration enables post-selection: by discarding low-confidence predictions (measured by the output confidence), we achieve lower logical error rates at the cost of reduced acceptance rate (\cref{fig:calibration}(a)).
While the waterfall effect reduces the number of qubits needed for a given target error rate, calibrated post-selection reduces the \emph{time} overhead.
On the $\llbracket 72, 12, 6 \rrbracket$ code at high noise level  $p=0.55\%$, our decoder reaches an error rate $\sim 2 \times 10^{-3}$ with an acceptance rate of ${\sim}95\%$, compared to ${\sim}5\%$ for cluster-based post-selection methods~\cite{lee2025efficient}---a roughly $20\times$ reduction in the number of retries required by repeat-until-success protocols~\cite{smith2024mitigating,zhou2025error,menon2025magic}. In~\cref{fig:calibration}(b), we show that even a modest 0.5\% per-cycle discard rate can yield up to two orders-of-magnitude reduction in $P_L$.
Such confidence-aware decoding could reduce the overhead of distillation protocols---including magic state distillation~\cite{campbell2017roads} and entanglement distillation~\cite{zhou2025error}---that rely on post-selected operations.

\subsection*{Discussion and outlook}
\label{sec:discussion}

These results have several implications for fault-tolerant quantum computation.
First, code design and resource estimation should move beyond distance as the sole figure of merit.
The sparsity of low-weight uncorrectable errors, the weight distribution of logical operators, and decoder accuracy all shape effective error suppression and can differ dramatically between codes with identical $\llbracket n, k, d \rrbracket$ parameters.
Code-specific models that capture these structural properties will yield more favorable resource estimates than the standard distance-based formula.

Second, the decoder should be treated as an integral component of fault-tolerant architecture co-design rather than an independent subsystem.
The capacity scaling results (\cref{fig:emergence}) reveal a sharp threshold in expressive capacity: below it, decoders cannot represent the complex error patterns that dominate at moderate noise; above it, near-optimal performance emerges.
This explains why existing decoders miss the waterfall and underscores that decoder power directly determines how much of a code's error-correcting capability is realized in practice.

Third, our decoder's geometric inductive bias (locality, translation equivariance, and anisotropy) makes it a general framework, not a code-specific solution.
The same architecture achieves near-optimal accuracy on both surface codes and bivariate bicycle codes, code families with fundamentally different connectivity and encoding rates that have historically required entirely different decoding strategies~\cite{bausch2024learning,alphaqubit2,higgott2025sparse,roffe2020decoding,maurer2025relay,fowler2013optimal}.
Since these inductive biases apply to any code whose checks share a common local structure, we expect Cascade to extend to other high-rate qLDPC code families, including lifted-product codes~\cite{xu2024constant}, Kasai codes~\cite{kasai2026breaking}, and other codes with analogous geometric regularity~\cite{panteleev2021degenerate}, where the waterfall effect should be even more pronounced at larger code sizes.
The convolutional structure is also inherently hardware-friendly: its sparse, local connectivity, feed-forward computation, and deterministic latency enable spatial mapping onto FPGAs and ASICs, with roofline estimates suggesting lightweight variants could approach the ${\sim}1\,\mu$s budget for superconducting qubits (Supplementary Information).

Finally, we note that these results are particularly timely since  trapped-ion~\cite{ballance2016high}, superconducting~\cite{li2024realization}, and neutral-atom~\cite{evered2025highfidelity} systems have all reached entangling error rates around ${\sim}0.1\%$, placing current hardware systems in the regime where the waterfall effect can enable logical error rates sufficient for the realization of practical fault-tolerant quantum algorithms. 

\bibliographystyle{naturemag_arxiv.bst}
\bibliography{refs}

\setcounter{figure}{0}
\newcounter{EDfig}
\renewcommand{\figurename}{Extended Data Fig.}

\clearpage
\newpage


\section*{Methods}

\naturepar{Stabilizer codes and the decoding problem}
A stabilizer code encodes $k$ logical qubits into $n$ physical qubits, protecting against errors up to weight $\lfloor (d-1)/2 \rfloor$, where $d$ is the code distance~\cite{gottesman1997stabilizer}.
The code is defined by a set of commuting Pauli operators called stabilizers, which are measured repeatedly to detect errors.
In a \emph{memory experiment}, a logical state is initialized, stabilizer measurements are performed for $R$ rounds, and the logical state is then measured.
Errors occurring during this process---on gates, idle qubits, and measurements---cause stabilizer outcomes to change between rounds.
These changes, called \emph{detection events}, form the \emph{syndrome} that the decoder must interpret.

The decoding problem is to infer, from the syndrome, whether a logical error has occurred.
Because many different physical error patterns can produce the same syndrome (degeneracy), the decoder need only identify the \emph{equivalence class} of the error, not the exact error.
The key metric is the \emph{logical error rate} $P_L$: the probability that the decoder's correction, combined with the actual errors, results in a logical bit flip.
Throughout this paper, we report $P_L$ per cycle per logical qubit, extracted from the block error rate $P_{\mathrm{block}}$ over the full syndrome volume as
\begin{equation}
    P_L = \frac{1 - \bigl(2(1-P_{\mathrm{block}})^{1/k}-1\bigr)^{1/R}}{2},
    \label{eq:pl}
\end{equation}
where $R$ is the number of rounds and $k$ is the number of logical qubits~\footnote{$P_L$ is defined by inverting $P_{\mathrm{block}} = 1 - \bigl(\frac{1+(1-2P_L)^R}{2}\bigr)^k$, the block error rate of $k$ independent qubits each failing with odd-parity probability $[1-(1-2P_L)^R]/2$ over $R$ rounds.}.
Below the code's threshold error rate, $P_L$ decreases exponentially with code distance, characterized by the \emph{error suppression factor} $\Lambda > 1$ such that $P_L \sim \Lambda^{-\lfloor(d+1)/2\rfloor}$.

We focus on two code families with geometric structure.
The surface code arranges data qubits on a 2D grid with $X$-type and $Z$-type stabilizers on alternating faces~\cite{fowler2012surface}.
A distance-$d$ surface code uses $d^2$ data qubits to encode one logical qubit.
With $R$ rounds of syndrome measurement, the syndrome forms a 3D structure: two spatial dimensions plus time.
Crucially, the surface code exhibits \emph{spatiotemporal translation symmetry in the bulk}: the local structure around any stabilizer is identical up to a spatial shift.

Bivariate bicycle (BB) codes are quantum LDPC codes constructed on a torus $\mathbb{Z}_\ell \times \mathbb{Z}_m$~\cite{bravyi2024high}.
An $\llbracket n, k, d \rrbracket$ BB code encodes $k$ logical qubits into $n$ physical qubits with distance $d$.
For example, the $\llbracket 144, 12, 12 \rrbracket$ code (with $\ell = 12$, $m = 6$) encodes 12 logical qubits---a significantly higher rate than surface codes.
Like surface codes, BB codes have $X$-type and $Z$-type stabilizers, but with a different connectivity pattern defined by polynomials over the torus.
The key property for our purposes is that BB codes also exhibit translation symmetry: all $X$-stabilizers (and separately, all $Z$-stabilizers) have identical local neighborhoods up to translation on the torus.
This regularity enables weight-sharing across the code, just as for surface codes.

\naturepar{Prior decoding algorithms}
For surface codes, minimum-weight perfect matching (MWPM) is the dominant decoding approach~\cite{dennis2002topological,higgott2022pymatching,higgott2025sparse}.
Variants such as correlated matching~\cite{fowler2013optimal} and belief matching~\cite{higgott2023beliefmatching} incorporate correlations between error types, improving accuracy at modest computational cost.
Union-find decoders offer a faster alternative with simpler operations, though with somewhat lower accuracy~\cite{delfosse2021almost}.

For quantum LDPC codes, matching-based decoders do not directly apply due to the hypergraph structure.
Belief propagation (BP) is a natural choice but often fails on quantum codes: under BP's fixed update rules, quantum degeneracy induces \emph{symmetric trapping sets}~\cite{poulin2008iterative,raveendran2023trapping}, causing messages to oscillate or converge to incorrect solutions.
BP with ordered statistics decoding (BP+OSD) follows unconverged BP with Gaussian elimination~\cite{panteleev2021degenerate,roffe2020decoding}, improving accuracy at significant computational cost.
Variants such as localized statistics decoding (BP+LSD) reduce this cost somewhat~\cite{hillmann2024localized}.
Relay~\cite{muller2024relax} introduces learnable scalar memory terms that break trapping-set symmetries, similar to neural BP approaches for classical LDPC codes~\cite{nachmani2018deep}.
The success of these methods demonstrates that the problem is algorithmic, arising from BP's rigid update rules, rather than fundamental to degeneracy itself.
Graph neural network decoders~\cite{gong2023graphneuralnetworksenhanced,ninkovic2024decodingquantumldpccodes} replace BP's update rules with learned functions but treat all neighbors symmetrically rather than exploiting geometric regularity.
AlphaQubit~\cite{bausch2024learning,alphaqubit2} uses a transformer architecture with global attention, achieving near-optimal accuracy on surface codes and colour codes but with higher training cost and without application to high-rate quantum LDPC codes.

\naturepar{Geometric structure and convolution}
QEC codes exhibit three geometric regularities that provide a natural guide for decoder design: translation equivariance, locality, and anisotropy.

Translation equivariance follows from the regularity of the code. Consider the bivariate bicycle codes, which are defined on a torus $\mathbb{Z}_\ell \times \mathbb{Z}_m$ with stabilizers arranged in a periodic lattice.
Since every stabilizer has the same local connectivity, a spatial shift of the syndrome by $(\Delta\ell, \Delta m)$ shifts the posterior distribution over data-qubit errors correspondingly.
A translation-equivariant decoder respects this symmetry, producing correspondingly shifted outputs from shifted inputs.
For surface codes, boundaries break this symmetry locally, but the bulk of the decoding problem remains translation-invariant.

Locality is less obvious. Errors manifest as local syndrome patterns, but the decoding problem itself is global: the decoder must determine the equivalence class of the error, which depends on the error's topology across the entire code.
What makes locality viable is a natural hierarchy of scales in the error structure.
Small-scale errors produce local syndrome patterns that can be identified from nearby information alone; once these are resolved, the remaining error landscape is effectively coarser.
Successive layers of local operations thus implement a form of coarse-graining~\cite{duclos2010fast}, where each layer absorbs errors at its characteristic scale.
After $L \sim d$ layers, the receptive field spans the full code distance, and only the global topological ambiguity remains to be resolved.

Finally, anisotropy: the decoder should learn to transmit messages differently in different directions.
In the surface code's checkerboard structure, a stabilizer's horizontal neighbor is of opposite type (X$\leftrightarrow$Z) while its diagonal neighbor is of the same type---information arriving from these directions has fundamentally different meaning.
The temporal direction carries distinct structure as well: measurement errors produce isolated detection events in time, while data qubit errors produce correlated pairs across consecutive rounds.

Existing decoders already satisfy some of these properties with fixed update rules: belief propagation performs local message passing on the bipartite Tanner graph, and the cellular automaton decoder~\cite{lake2025local} adds anisotropy, labeling messages by coordinate direction to distinguish between different spatial neighbors---indeed, Ref.~\cite{lake2025local} proves that these three structural priors are fully compatible with fault-tolerant decoding of the surface code.
The remaining question is how to make the update rules learned while preserving all three properties.
Consider a graph neural network on the check graph, where nodes are stabilizers connected when their supports overlap.
At each layer, every node aggregates messages from its neighbors and updates its state.
A standard GNN treats all neighbors symmetrically (permutation equivariance), losing the directional structure that anisotropy requires.
A relational GNN~\cite{Schlichtkrull2018rgcn} fixes this by assigning different weight matrices to different edge types.
For codes with translation symmetry, the geometric offset between two stabilizers provides a natural edge typing, since the relation ``neighbor at offset $\delta$'' is consistently defined across the code.
Adding weight sharing across positions---the same learned rules at every site---then gives a convolution: learned, local, directional message passing with translation equivariance.
This is the same structure as the cellular automaton decoder, but with learned update rules operating in a high-dimensional embedding space rather than hand-designed rules in probability space.

\begin{figure*}
    \centering
    \begin{minipage}[c]{0.48\textwidth}
        \centering
        \begin{tikzpicture}[scale=0.4]
            \definecolor{col1}{RGB}{228,26,28}
            \definecolor{col2}{RGB}{55,126,184}
            \definecolor{col3}{RGB}{77,175,74}
            \definecolor{col4}{RGB}{152,78,163}
            \definecolor{col5}{RGB}{255,127,0}
            \definecolor{col6}{RGB}{166,86,40}
            \definecolor{col7}{RGB}{247,129,191}
            \definecolor{col8}{RGB}{153,153,153}
            \begin{scope}[xshift=0cm, yshift=0cm]
                \node[anchor=north] at (3, 7.5) {\textbf{Convolution}};
                \foreach \x in {0,...,6} {
                    \foreach \y in {0,...,6} {
                        \fill[black!10] (\x, \y) circle (0.15);
                    }
                }
                \node[circle, fill=black, inner sep=2pt] (c1) at (2, 5) {};
                \foreach \dx/\dy/\col in {-1/1/col1, 0/1/col2, 1/1/col3, -1/0/col4, 1/0/col5, -1/-1/col6, 0/-1/col7, 1/-1/col8} {
                    \draw[->, \col, line width=1.5pt] (c1) -- +(\dx, \dy);
                }
                \node[circle, fill=black, inner sep=2pt] (c2) at (4, 2) {};
                \foreach \dx/\dy/\col in {-1/1/col1, 0/1/col2, 1/1/col3, -1/0/col4, 1/0/col5, -1/-1/col6, 0/-1/col7, 1/-1/col8} {
                    \draw[->, \col, line width=1.5pt] (c2) -- +(\dx, \dy);
                }
            \end{scope}
            \begin{scope}[xshift=10.5cm, yshift=0cm]
                \node[anchor=north] at (3, 7.5) {\textbf{Local Attention}};
                \foreach \x in {0,...,6} {
                    \foreach \y in {0,...,6} {
                        \fill[black!10] (\x, \y) circle (0.15);
                    }
                }
                \node[circle, fill=black, inner sep=2pt] (l1) at (2, 5) {};
                \foreach \dx/\dy/\thickness in {-1/1/0.5, 0/1/2.5, 1/1/1.0, -1/0/1.5, 1/0/3.0, -1/-1/0.8, 0/-1/1.2, 1/-1/2.0} {
                    \draw[->, black!60, line width=\thickness pt] (l1) -- +(\dx, \dy);
                }
                \node[circle, fill=black, inner sep=2pt] (l2) at (4, 2) {};
                \foreach \dx/\dy/\thickness in {-1/1/2.5, 0/1/0.8, 1/1/1.8, -1/0/0.5, 1/0/1.0, -1/-1/2.8, 0/-1/1.5, 1/-1/0.6} {
                    \draw[->, black!60, line width=\thickness pt] (l2) -- +(\dx, \dy);
                }
            \end{scope}
            \begin{scope}[xshift=5.25cm, yshift=-8cm]
                \node[anchor=north] at (3, 7.5) {\textbf{Full Attention}};
                \foreach \x in {0,...,6} {
                    \foreach \y in {0,...,6} {
                        \fill[black!10] (\x, \y) circle (0.15);
                    }
                }
                \node[circle, fill=black, inner sep=2pt] (f1) at (3, 3) {};
                \foreach \x/\y/\thickness in {0/0/0.5, 2/0/1.5, 4/0/2.0, 6/0/0.8, 6/1/2.5, 0/2/1.0, 6/3/0.6, 0/4/1.8, 6/5/1.3, 0/6/0.9, 6/6/2.2, 1/6/1.1, 2/6/1.6, 4/6/0.7, 5/6/1.9, 1/5/0.8, 2/1/1.7, 5/1/1.0, 1/3/2.3, 5/3/0.9, 2/5/1.5, 5/5/1.2, 1/4/2.0} {
                    \draw[->, black!40, line width=\thickness pt] (f1) -- (\x, \y);
                }
            \end{scope}
        \end{tikzpicture}
    \end{minipage}
    \hfill
    \begin{minipage}[c]{0.48\textwidth}
        \centering
        \includegraphics[width=\linewidth]{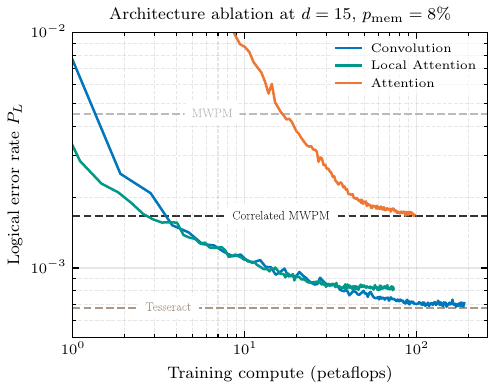}
    \end{minipage}
    \caption{\textbf{Architectural ablation on surface codes.}
    (a) Architectural inductive biases visualized on 2D syndrome grids.
    \textbf{Convolution}: Arrows from two different positions show identical directional patterns (colors encode direction-specific learned weights), demonstrating translation equivariance combined with learned anisotropy.
    \textbf{Local attention}: Arrow thickness represents learned attention weights; different thickness patterns at different positions show the architecture is position-dependent rather than translation-equivariant.
    \textbf{Full attention}: Global all-to-all connectivity with position-dependent learned weights.
    (b) Logical error rate versus training compute (PFLOPs) for three architectures with fixed depth ($L=8$) and width ($H=256$) at distance $d=15$, evaluated at physical error rate $p=8\%$. Convolution achieves the lowest error rate, while local attention and full attention saturate at higher error rates despite equal or greater training compute. Existing decoders (MWPM, Correlated MWPM, Tesseract) shown for reference.}
    \refstepcounter{EDfig}\label{fig:ablation}
\end{figure*}

A natural question is whether attention-based architectures---which have achieved state-of-the-art results on surface codes~\cite{alphaqubit2}---might learn faster or more effectively than convolutions when training compute is held fixed.
Using the same setting as the scaling study (data-level noise, $d=15$), we compare three architectures with identical model size ($L=8$ layers, $H=256$ hidden dimension): standard convolution, local (neighborhood) attention~\cite{hassani2023neighborhood}, and full global attention.
To ensure a fair comparison across architectures with different per-step computational costs, we plot logical error rate as a function of training compute rather than training steps.
The results (Extended Data Fig.~\ref{fig:ablation}b) show that convolution achieves the lowest final error rate, reaching parity with the Tesseract decoder, while local attention saturates at a higher error rate.
Full attention performs worst: not only does it saturate at an even higher error rate than local attention, but its entire learning curve is shifted rightward, requiring more total training compute to reach any given accuracy level.

These results decompose the contribution of the three geometric properties (Extended Data Fig.~\ref{fig:ablation}a).
The largest gain comes from locality: restricting attention to local neighborhoods (full $\to$ local attention) substantially improves both final accuracy and compute efficiency.
The further improvement from local attention to convolution reflects the addition of translation equivariance and anisotropy---convolutions apply identical direction-specific rules at every position, whereas local attention learns position-dependent weights that treat neighbors symmetrically.
That adding flexibility (local $\to$ global attention) degrades rather than helps performance confirms that these structural priors are not merely convenient but essential.

The specific form of the convolution depends on the code.
For surface codes, the syndrome is a 3D tensor of shape $(R, d+1, d+1)$ (time plus two spatial dimensions), and the convolution is a standard 3D convolution with zero padding at boundaries.
For BB codes on a torus $\mathbb{Z}_\ell \times \mathbb{Z}_m$, the spatial neighborhood of each stabilizer is defined by the Tanner graph rather than a grid stencil, while the temporal direction uses a standard 1D convolution over adjacent time steps. The update for hidden state $h_v$ at position $v$ is
\begin{equation}
    h_v^{(l+1)} = \sum_{u \in \mathcal{N}(v)} W_{\delta(u,v)}^{(l)} h_u^{(l)} + W_{\text{self}}^{(l)} h_v^{(l)},
\end{equation}
where $\mathcal{N}(v)$ is the full spatiotemporal neighborhood of $v$ and $\delta(u,v)$ is the relative position of $u$ with respect to $v$ (spatial offset on the torus and temporal offset).
The weight matrix $W_{\delta}^{(l)}$ depends only on the relative position $\delta$, not on the absolute position $v$.

\naturepar{Architecture}
Our architecture follows a standard encoder-readout structure, with the convolutional operator as the only code-specific component.
The input is the syndrome: a tensor of binary detection events indicating where stabilizer measurements changed between rounds.
We embed each detection event into a hidden representation of dimension $H$ using a learned embedding layer.

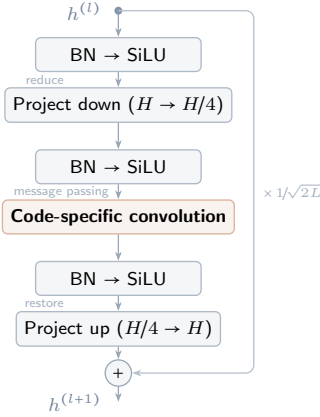
\begin{figure*}
\centering
\begin{tikzpicture}[
    op/.style={draw=deepslate!40, fill=deepslate!5, rounded corners=2pt,
        minimum width=2.2cm, minimum height=0.45cm,
        font=\sffamily\scriptsize, align=center, inner sep=3pt, line width=0.5pt},
    convop/.style={draw=warmrust!50, fill=warmrust!8, rounded corners=2pt,
        minimum width=2.2cm, minimum height=0.45cm,
        font=\sffamily\scriptsize\bfseries, align=center, inner sep=3pt, line width=0.5pt},
    arr/.style={-{Stealth[length=3.5pt, width=2.5pt]}, line width=0.5pt, deepslate!50},
    skipline/.style={line width=0.5pt, deepslate!35, rounded corners=3pt},
    stagelab/.style={font=\sffamily\tiny, text=deepslate!45},
]
\node[op] (bn1) {BN $\to$ SiLU};
\node[op, below=0.2cm of bn1] (proj1) {Project down ($H \to H\!/4$)};
\node[op, below=0.35cm of proj1] (bn2) {BN $\to$ SiLU};
\node[convop, below=0.2cm of bn2] (conv) {Code-specific convolution};
\node[op, below=0.35cm of conv] (bn3) {BN $\to$ SiLU};
\node[op, below=0.2cm of bn3] (proj2) {Project up ($H\!/4 \to H$)};
\draw[arr] (bn1) -- (proj1);
\draw[arr] (proj1) -- (bn2);
\draw[arr] (bn2) -- (conv);
\draw[arr] (conv) -- (bn3);
\draw[arr] (bn3) -- (proj2);
\coordinate (fork) at ($(bn1.north)+(0, 0.35)$);
\node[font=\sffamily\scriptsize, deepslate!70, left=0.1cm of fork] {$h^{(l)}$};
\draw[arr] (fork) -- (bn1.north);
\fill[deepslate!60] (fork) circle (1.5pt);
\coordinate (sumpos) at ($(proj2.south)+(0, -0.35)$);
\node[circle, draw=deepslate!50, fill=deepslate!5, inner sep=1pt,
    font=\sffamily\scriptsize\bfseries, line width=0.5pt, minimum size=0.35cm] (sum) at (sumpos) {$+$};
\draw[arr] (proj2) -- (sum);
\draw[skipline, -{Stealth[length=3.5pt, width=2.5pt]}]
    (fork) -- ++(1.8, 0) |- node[pos=0.25, right, stagelab] {$\times\, 1\!/\!\sqrt{2L}$} (sum.east);
\coordinate (out) at ($(sum.south)+(0, -0.2)$);
\draw[arr] (sum) -- (out);
\node[font=\sffamily\scriptsize, deepslate!70, left=0.1cm of out] {$h^{(l+1)}$};
\node[stagelab, left=0.6cm of $(bn1)!0.5!(proj1)$] {reduce};
\node[stagelab, left=0.cm of $(bn2)!0.5!(conv)$] {message passing};
\node[stagelab, left=0.6cm of $(bn3)!0.5!(proj2)$] {restore};
\end{tikzpicture}
\caption{\textbf{Bottleneck residual block.}
The network backbone consists of $L$ identically-structured blocks with independent learned parameters, each processing a hidden representation of dimension $H$ at every syndrome location.
First, the representation is projected from $H$ to $H/4$ dimensions (reduce); the code-specific convolution then operates in this lower-dimensional space (message passing), after which the representation is projected back to $H$ dimensions (restore).
The bottleneck reduces the cost of the convolution by roughly $16\times$ while preserving expressive capacity.
Each projection and convolution is preceded by batch normalization (BN), which standardizes activations to zero mean and unit variance, and a SiLU nonlinearity.
A scaled residual connection adds the block input $h^{(l)}$ directly to the output, weighted by $1/\sqrt{2L}$, so that each block learns a correction to the identity mapping rather than reconstructing the full representation---a standard technique for stable training of deep networks.}
\refstepcounter{EDfig}\label{fig:block}
\end{figure*}

The backbone consists of $L$ identically-structured processing blocks stacked sequentially, each with independent learned parameters, following a bottleneck residual design~\cite{kaiming2016identity} (Extended Data Fig.~\ref{fig:block}).
In each block, the $H$-dimensional representation is first projected down to $H/4$ dimensions, processed by the code-specific convolution, and then projected back to $H$ dimensions.
Because the expensive convolution operates in the lower-dimensional space, this reduces its cost by roughly $16\times$ while preserving expressive capacity.
Each projection and convolution is preceded by batch normalization (which standardizes activations to zero mean and unit variance, stabilizing training) and a SiLU nonlinearity (a smooth activation function that enables the network to learn nonlinear relationships).
A residual connection adds each block's input directly to its output, so the block only needs to learn a correction to the identity mapping rather than reconstructing the full representation---a standard technique that enables stable training of deep networks.
The hidden dimension $H$ is constant throughout the backbone.

The code-specific convolution operates on the check graph, where the network maintains hidden state at each syndrome location.
Two checks are neighbors if and only if they share a data qubit; for codes with high-weight stabilizers, this neighborhood can be large.
Any such check-to-check convolution can be factored into two bipartite steps---check $\to$ data $\to$ check---each with a smaller kernel defined by the Tanner graph's local connectivity.
This admits a natural interpretation as message passing on the bipartite Tanner graph, and reduces the number of learned relations per step while preserving the same receptive field per layer.
For BB codes, the savings are substantial: each check has 22 spatial neighbors in the check-to-check graph across 3 temporal offsets, giving 66 distinct relations per layer, while each bipartite step involves only 6 spatial neighbors across 2 temporal offsets (12 relations)---an over $5\times$ reduction in kernel size.
We implement these bipartite convolutions using custom Triton kernels that exploit the regular structure for efficient memory access.
For surface codes, the direct $3 \times 3 \times 3$ kernel is already small, and factorization offers insufficient savings to justify the overhead.
A further reduction is possible via depthwise convolution~\cite{howard2017mobilenets,liu2022convnet}, which processes each channel independently in the spatial operation, reducing the spatial cost from $O(K(H/b)^2)$ to $O(KH/b)$ per layer.

A key architectural consideration is the relationship between network depth $L$ and code distance $d$.
For surface codes, error correlations extend over spatial scales $\sim d$, requiring the network's receptive field to grow with distance.
With typical $3\times 3\times 3$ convolutional kernels, each layer increases the receptive field by 2 in space and time, giving a receptive field of $\sim 2L$ after $L$ layers.
To ensure sufficient spatial context for decoding, we scale depth with distance, using $L \sim d$ for our experiments.

After the final convolutional block, a convolution scatters the check-node representations to data qubits. We then aggregate information for each logical observable by average pooling over the data qubits in that observable's support.
The pooled representation is passed through a two-layer multilayer perceptron (with hidden dimension $2H$) to produce a logit for each logical observable.
For surface codes, which encode a single logical qubit, the output is a single logit predicting whether an $X_L$ or $Z_L$ error occurred.
For BB codes, which encode multiple logical qubits, the output is a vector of logits, one per logical observable.

The architecture is parameterized by depth $L$ and width $H$.
We train a separate model for each $(H, L)$ configuration.
To ensure stable training across different widths, we use Maximal Update Parameterization (MuP)~\cite{yang2022tensor}, which adjusts learning rates and initialization scales as a function of width so that a single set of training hyperparameters works across all model sizes without per-configuration tuning.

\naturepar{Training}
We generate training data using Stim~\cite{gidney2021stim}, simulating memory experiments with $R = d$ rounds of syndrome extraction under circuit-level depolarizing noise.
Each training example consists of a syndrome (the detection events across all $R$ rounds) and a label (whether a logical error occurred).
Data is generated on-the-fly during training, providing an effectively unlimited dataset; in practice, our models converge within $3 \times 10^8$ training examples.
We train with binary cross-entropy loss on the logical error prediction.
For codes with multiple logical observables, we average the cross-entropy losses across all observables.

We use the Muon optimizer~\cite{jordan2024muon}, which applies Newton-Schulz orthogonalization to gradient updates for matrix-valued parameters (convolution weights), with Lion~\cite{chen2023symbolic} for scalar parameters (biases, normalization parameters, embeddings, and readout layers).
Peak learning rates are $3 \times 10^{-3}$ for Muon and $2 \times 10^{-4}$ for Lion, following a cosine schedule over $50000$ steps (decaying to $\frac{1}{10}$ of peak) with linear warmup over $1000$ steps.
We use a weight decay of $3 \times 10^{-3}$ and maintain an exponential moving average (EMA) of model weights~\cite{pavel2018averaging} with decay rate $\beta=0.9998$.
All reported results use the EMA weights.
We train in mixed precision (bfloat16) with gradient norm clipping, a batch size of $3328$, for $80000$ steps. The largest model we train (for the $d=19$ surface code decoder) converges in approximately 200 H200 GPU hours; the largest BB code model converges in under 100.

While we ultimately want to train at high physical error rates (where gradient signal is plentiful), we find empirically that training from random initialization directly at high $p$ leads to prolonged periods where the network fails to learn better than random---a phenomenon reminiscent of ``grokking''~\cite{power2022grokking,liu2022towards}.
We address this with a simple three-stage curriculum (similar to the one used in~\cite{maskara2019advantages}) that bootstraps the network from easier to harder problems.
First, we train briefly at a low noise level $p_1$ where errors are sparse and the network quickly achieves better-than-random performance.
Second, we linearly anneal the noise level from $p_1$ to the target $p_2$ over a fixed number of steps.
Third, we continue training at the target noise level $p_2$ until convergence.
The first two stages are brief, accounting for at most 2\% of the total training steps; the vast majority of training occurs at the target noise level.

We train at a single, relatively high physical error rate---$p = 0.7\%$ for surface codes and $p = 0.55\%$ for BB codes under circuit-level noise---and evaluate across a range of error rates extending to much lower values, avoiding the cost of generating rare low-error-rate samples.
For the scaling study (\cref{fig:emergence}), we trained surface code decoders of varying width $H$ with fixed depth $L=8$ under data-level depolarizing noise at $p = 13\%$; we use data-level noise for this experiment because it enables faster training of the many model configurations needed for a systematic study.
The error suppression exponent $m$ (from $P_L \propto p^m$) saturates at $m \approx 8$ for $H \gtrsim 64$; beyond this point, further increases in width continue to reduce logical error rates through the prefactor while the exponent remains roughly constant.

\naturepar{Data Availability}
The data that supports the findings of this study are available from the corresponding author on request. \\

\noindent\textbf{Acknowledgments}
We acknowledge helpful discussions with Varun Menon and Nazli Ugur Koyluoglu. We thank Hengyun (Harry) Zhou, Nishad Maskara, and Johannes Bausch for providing feedback on earlier versions of this manuscript. A.G. and J.P.B.A. acknowledge support from the Unitary Foundation. The computations in this paper were run on the FASRC Cannon cluster supported by the FAS Division of Science Research Computing Group at Harvard University. This research used resources of the National Energy Research Scientific Computing Center (NERSC), a Department of Energy Office of Science User Facility under Contract No. DE-AC02-05CH11231 using NERSC award DDR-ERCAP0038713. We acknowledge support from the National Science Foundation  (PHY-2012023 and CCF-2313084) and  through  the CUA Physics Frontiers Center (PHY-2317134) and NVQL (PHY-2410716) and from DOE through the QUACQ collaboration (DE-SC0025572), Quantum Systems Accelerator Center (DE-AC02-05CH11231), IARPA and the Army Research Office, under the Entangled Logical Qubits program (W911NF-23-2-0219), the DARPA MeasQuIT program (HR0011-24-9-0359). \\

\noindent\textbf{Author contributions}
A.G. and J.P.B.A. conceived the method. A.G. implemented and  performed numerical simulations. M.D.L. and S.F.Y. supervised the project. All authors discussed the results and contributed to  the manuscript. \\

\noindent\textbf{Competing interests:} M.D.L. is a co-founder, shareholder and Chief Scientist of QuEra Computing. S.F.Y. is a spouse of a shareholder of QuEra Computing. A.G. and J.P.B.A. have served as consultants for QuEra Computing. 
\\

\noindent\textbf{Correspondence and requests for materials} should be addressed to M.D.L and S.F.Y. \\


\clearpage
\newpage

\section*{Supplementary Information: Toward real-time decoding on dedicated hardware}
\label{sec:hardware-appendix}

The main text argues that convolutional decoders are well-suited to hardware implementation due to their local connectivity, regular computation patterns, and translation-equivariant structure.
This section provides supporting analysis, combining one experimentally validated result---post-training quantization to FP8 with no accuracy loss---with projections and design considerations that have not yet been implemented in hardware: a computational cost comparison of architectural variants, roofline latency estimates, and a dataflow architecture outline.

\naturepar{Quantization and inference optimization}
\label{sec:quantization}
Deploying neural decoders on FPGA or ASIC hardware requires low-precision arithmetic: the area and power consumption of a multiply-accumulate (MAC) unit---which computes $a \leftarrow a + b \times c$ in a single operation---scale roughly quadratically with operand bit width, so reducing precision from FP32 to FP8 yields an approximately $16\times$ reduction in multiplier area and power.
We use MACs rather than floating-point operations (FLOPs) as our complexity measure because a MAC is the atomic hardware primitive: each MAC corresponds to exactly one multiplier circuit, making MAC counts a direct proxy for chip area, power, and latency on dedicated hardware.

We find that our convolutional decoders tolerate aggressive post-training quantization with no measurable accuracy degradation.
At inference time, we first \emph{fold} batch normalization parameters into the preceding convolution weights: the affine transformation $y = \gamma \frac{x - \mu}{\sqrt{\sigma^2 + \epsilon}} + \beta$ is absorbed into modified weights $W_{\text{fold}} = \frac{\gamma}{\sqrt{\sigma^2 + \epsilon}} W$ and biases $b_{\text{fold}} = \gamma \frac{b - \mu}{\sqrt{\sigma^2 + \epsilon}} + \beta$.
This eliminates all normalization layers from the inference graph, leaving a pure sequence of convolutions, pointwise nonlinearities, global pooling, and a linear head.
We then use 8-bit floating point (FP8) inference with \texttt{torchao}~\cite{torchao}.
The resulting model achieves logical error rates indistinguishable from the full-precision model across all physical error rates tested, with no quantization-aware retraining required.

We attribute this robustness to properties of the convolutional architecture: (i) convolution weights have regular, well-behaved distributions without the outlier channels that complicate quantization of attention projections, (ii) the absence of softmax eliminates the extreme dynamic range that necessitates high precision in transformer architectures, and (iii) SiLU activations have smooth, well-behaved output distributions without the extreme values that arise in attention layers.
These properties make post-training quantization straightforward, in contrast to transformer-based decoders where quantization typically requires specialized techniques.

The combination of FP8 arithmetic with the depthwise convolution variant yields a compound reduction in hardware cost: $\sim 4\times$ from depthwise factorization (reducing spatial convolution FLOPs) and $\sim 16\times$ from FP8 versus FP32 precision, for an overall $\sim 60\times$ reduction in multiplier area relative to a standard FP32 convolution baseline.
Further reduction to INT4 via quantization-aware training is a natural direction for future work.

\naturepar{Computational cost and roofline latency}
\label{sec:flop-comparison}
\label{sec:roofline}
The architectural ablation in the main text compares convolution, local attention, and full attention at fixed model size.
Here we analyze the per-layer computational costs to contextualize the accuracy--compute trade-off and estimate achievable latencies on dedicated hardware.

\begin{figure*}
    \centering
    \begin{tikzpicture}[
        sp/.style={
            draw=warmrust!50, fill=warmrust!8,
            rounded corners=2pt, minimum height=0.45cm,
            font=\sffamily\scriptsize\bfseries, align=center,
            inner sep=3pt, line width=0.5pt,
        },
        arr/.style={-{Stealth[length=3.5pt, width=2.5pt]}, line width=0.5pt, deepslate!50},
        skipline/.style={line width=0.5pt, deepslate!35, rounded corners=3pt},
        coltitle/.style={font=\sffamily\small\bfseries, text=deepslate!80, align=center},
        dimlab/.style={font=\sffamily\tiny, text=deepslate!45},
    ]
    \def\colsep{3.6}
    \def\vsep{0.55}
    \def\boxh{0.45}
    \def\halfh{0.225}
    \pgfmathsetmacro{\stride}{\boxh + \vsep}
    \def\hwn{0.55}
    \def\hwm{0.9}
    \def\hww{1.3}
    \def\wnarrow{1.1}
    \def\wmed{1.8}
    \def\wwide{2.6}
    \begin{scope}[xshift=0cm]
        \node[coltitle] at (0, 1.0) {Bottleneck~{\hypersetup{citecolor=mauve}\cite{kaiming2016identity}}\\[-2pt]{\fontseries{m}\selectfont\tiny(Used in this work)}};
        \coordinate (a-in) at (0, 0.45);
        \node[dimlab, left=0.05cm of a-in] {$H$};
        \filldraw[fill=deepslate!5, draw=deepslate!40, rounded corners=2pt, line width=0.5pt]
            (-\hww, \halfh) -- (\hww, \halfh) -- (\hwm, -\halfh) -- (-\hwm, -\halfh) -- cycle;
        \node[font=\sffamily\scriptsize] at (0, 0) {$1\!\times\!1$: $H\!\to\!H\!/b$};
        \coordinate (a1-top) at (0, \halfh);
        \coordinate (a1-bot) at (0, -\halfh);
        \node[sp, minimum width=\wmed cm] (a2) at (0, {-\stride}) {Conv ($H\!/b$)};
        \pgfmathsetmacro{\yc}{-2*\stride}
        \filldraw[fill=deepslate!5, draw=deepslate!40, rounded corners=2pt, line width=0.5pt]
            (-\hwm, \yc+\halfh) -- (\hwm, \yc+\halfh) -- (\hww, \yc-\halfh) -- (-\hww, \yc-\halfh) -- cycle;
        \node[font=\sffamily\scriptsize] at (0, \yc) {$1\!\times\!1$: $H\!/b\!\to\!H$};
        \coordinate (a3-top) at (0, {\yc+\halfh});
        \coordinate (a3-bot) at (0, {\yc-\halfh});
        \draw[arr] (a-in) -- (a1-top);
        \draw[arr] (a1-bot) -- (a2.north);
        \draw[arr] (a2.south) -- (a3-top);
        \node[circle, draw=deepslate!50, fill=deepslate!5, inner sep=1pt,
            font=\sffamily\scriptsize\bfseries, line width=0.5pt, minimum size=0.3cm]
            (a-sum) at (0, {\yc-\halfh-0.35}) {$\!+\!$};
        \draw[arr] (a3-bot) -- (a-sum);
        \fill[deepslate!60] (a-in) circle (1.2pt);
        \draw[skipline, -{Stealth[length=3pt, width=2pt]}]
            (a-in) -- ++(\wwide/2+0.3, 0) |- (a-sum.east);
        \coordinate (a-out) at ($(a-sum.south)+(0, -0.15)$);
        \draw[arr] (a-sum) -- (a-out);
        \node[dimlab, left=0.05cm of a-out] {$H$};
    \end{scope}
    \begin{scope}[xshift=\colsep cm]
        \node[coltitle] at (0, 1.0) {ConvNeXt~{\hypersetup{citecolor=mauve}\cite{liu2022convnet}} \\[-2pt]{\fontseries{m}\selectfont\tiny}};
        \coordinate (c-in) at (0, 0.45);
        \node[dimlab, left=0.05cm of c-in] {$H$};
        \node[sp, minimum width=\wmed cm] (c1) at (0, 0) {DW Conv ($H$)};
        \pgfmathsetmacro{\yb}{-\stride}
        \filldraw[fill=deepslate!5, draw=deepslate!40, rounded corners=2pt, line width=0.5pt]
            (-\hwm, \yb+\halfh) -- (\hwm, \yb+\halfh) -- (\hww, \yb-\halfh) -- (-\hww, \yb-\halfh) -- cycle;
        \node[font=\sffamily\scriptsize] at (0, \yb) {$1\!\times\!1$: $H\!\to\!bH$};
        \coordinate (c2-top) at (0, {\yb+\halfh});
        \coordinate (c2-bot) at (0, {\yb-\halfh});
        \pgfmathsetmacro{\yc}{-2*\stride}
        \filldraw[fill=deepslate!5, draw=deepslate!40, rounded corners=2pt, line width=0.5pt]
            (-\hww, \yc+\halfh) -- (\hww, \yc+\halfh) -- (\hwm, \yc-\halfh) -- (-\hwm, \yc-\halfh) -- cycle;
        \node[font=\sffamily\scriptsize] at (0, \yc) {$1\!\times\!1$: $bH\!\to\!H$};
        \coordinate (c3-top) at (0, {\yc+\halfh});
        \coordinate (c3-bot) at (0, {\yc-\halfh});
        \draw[arr] (c-in) -- (c1.north);
        \draw[arr] (c1.south) -- (c2-top);
        \draw[arr] (c2-bot) -- (c3-top);
        \node[circle, draw=deepslate!50, fill=deepslate!5, inner sep=1pt,
            font=\sffamily\scriptsize\bfseries, line width=0.5pt, minimum size=0.3cm]
            (c-sum) at (0, {\yc-\halfh-0.35}) {$\!+\!$};
        \draw[arr] (c3-bot) -- (c-sum);
        \fill[deepslate!60] (c-in) circle (1.2pt);
        \draw[skipline, -{Stealth[length=3pt, width=2pt]}]
            (c-in) -- ++(\wwide/2+0.3, 0) |- (c-sum.east);
        \coordinate (c-out) at ($(c-sum.south)+(0, -0.15)$);
        \draw[arr] (c-sum) -- (c-out);
        \node[dimlab, left=0.05cm of c-out] {$H$};
    \end{scope}
    \begin{scope}[xshift=2*\colsep cm]
        \node[coltitle] at (0, 1.0) {Transformer~{\hypersetup{citecolor=mauve}\cite{vaswani2017attention}}\\[-2pt]{\fontseries{m}\selectfont\tiny (Used in~{\hypersetup{citecolor=mauve}\cite{bausch2024learning,ataides2025neural,alphaqubit2}})}};
        \coordinate (d-in) at (0, 0.45);
        \node[dimlab, left=0.05cm of d-in] {$H$};
        \node[sp, minimum width=\wmed cm] (d1) at (0, 0) {MHA ($H$)};
        \pgfmathsetmacro{\yb}{-\stride}
        \filldraw[fill=deepslate!5, draw=deepslate!40, rounded corners=2pt, line width=0.5pt]
            (-\hwm, \yb+\halfh) -- (\hwm, \yb+\halfh) -- (\hww, \yb-\halfh) -- (-\hww, \yb-\halfh) -- cycle;
        \node[font=\sffamily\scriptsize] at (0, \yb) {$1\!\times\!1$: $H\!\to\!bH$};
        \coordinate (d2-top) at (0, {\yb+\halfh});
        \coordinate (d2-bot) at (0, {\yb-\halfh});
        \pgfmathsetmacro{\yc}{-2*\stride}
        \filldraw[fill=deepslate!5, draw=deepslate!40, rounded corners=2pt, line width=0.5pt]
            (-\hww, \yc+\halfh) -- (\hww, \yc+\halfh) -- (\hwm, \yc-\halfh) -- (-\hwm, \yc-\halfh) -- cycle;
        \node[font=\sffamily\scriptsize] at (0, \yc) {$1\!\times\!1$: $bH\!\to\!H$};
        \coordinate (d3-top) at (0, {\yc+\halfh});
        \coordinate (d3-bot) at (0, {\yc-\halfh});
        \draw[arr] (d-in) -- (d1.north);
        \draw[arr] (d1.south) -- (d2-top);
        \draw[arr] (d2-bot) -- (d3-top);
        \node[circle, draw=deepslate!50, fill=deepslate!5, inner sep=1pt,
            font=\sffamily\scriptsize\bfseries, line width=0.5pt, minimum size=0.3cm]
            (d-sum) at (0, {\yc-\halfh-0.35}) {$\!+\!$};
        \draw[arr] (d3-bot) -- (d-sum);
        \fill[deepslate!60] (d-in) circle (1.2pt);
        \draw[skipline, -{Stealth[length=3pt, width=2pt]}]
            (d-in) -- ++(\wwide/2+0.3, 0) |- (d-sum.east);
        \coordinate (d-out) at ($(d-sum.south)+(0, -0.15)$);
        \draw[arr] (d-sum) -- (d-out);
        \node[dimlab, left=0.05cm of d-out] {$H$};
    \end{scope}
    \pgfmathsetmacro{\legy}{-2*\stride - \halfh - 1.25}
    \filldraw[fill=deepslate!5, draw=deepslate!40, rounded corners=2pt, line width=0.5pt]
        ({0.5*\colsep - 0.4}, \legy) rectangle ++({0.8}, 0.25);
    \node[font=\sffamily\tiny, text=deepslate!60] at ({0.5*\colsep + 0.85}, {\legy + 0.125}) {Pointwise};
    \filldraw[fill=warmrust!8, draw=warmrust!50, rounded corners=2pt, line width=0.5pt]
        ({1.5*\colsep - 0.4}, \legy) rectangle ++({0.8}, 0.25);
    \node[font=\sffamily\tiny, text=deepslate!60] at ({1.5*\colsep + 0.85}, {\legy + 0.125}) {Spatial};
    \end{tikzpicture}
    \caption{\textbf{Block decomposition across architectures.}
    Every block decomposes into two pointwise projections (grey) and one spatial operation (orange), differing only in their ordering and the dimension at which each operates.
    Trapezoids indicate dimension changes; rectangles indicate same-dimension operations; shape width encodes the channel dimension at each stage.
    The bottleneck contracts to $H/b$ before the spatial step; ConvNeXt and transformer blocks apply the spatial operation at full width $H$ followed by an expand--contract MLP.
    Under the relabeling $H \to Hb$, all three variants yield the same MAC cost structure (\cref{eq:macs}).
    Pointwise nonlinearities and normalizations are omitted from the diagram as their cost is negligible compared to the linear projections and spatial operations.}
    \refstepcounter{EDfig}\label{fig:block-comparison}
\end{figure*}
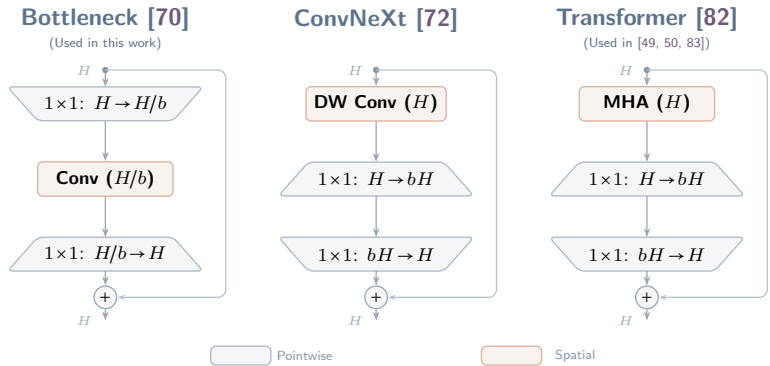

Each bottleneck block with hidden dimension $H$, bottleneck factor $b$, kernel size $K$, and $n$ syndrome positions consists of two pointwise projections ($H \to H/b$ and $H/b \to H$) plus a spatial operation.
We measure cost in multiply-accumulate operations (MACs), since each MAC maps to exactly one multiplier circuit on dedicated hardware, making it a direct proxy for chip area, power, and latency.
The per-block MAC counts for each architectural variant are:
\begin{equation}
\begin{aligned}
    \text{Convolution:} \quad & \sim 2nH^2/b + nK (H/b)^2 \\
    \text{Depthwise conv.:} \quad & \sim 2nH^2/b + nK (H/b) \\
    \text{Local attention:} \quad & \sim 2nH^2/b + nK (H/b) \\
    & \quad + 3n(H/b)^2 \\
    \text{Full attention:} \quad & \sim 2nH^2/b + n^2 (H/b) \\
    & \quad + 3n(H/b)^2
\end{aligned}    
\label{eq:macs}
\end{equation}
In each case, the first term is the shared pointwise projection cost; the remaining terms are the spatial operation, which varies across architectures.
Both convolution and local attention scale linearly in $n$; full attention scales quadratically.
These decompositions are not specific to our bottleneck design:
ConvNeXt blocks~\cite{liu2022convnet} and transformer blocks~\cite{vaswani2017attention,alphaqubit2} admit the same decomposition into two pointwise projections and one spatial operation, yielding identical MAC formulas under the relabeling $H \to Hb$ (Extended Data Fig.~\ref{fig:block-comparison}).

For a network of depth $L$, the total MACs per round is $L$ times the per-block cost.
At representative parameters ($H = 256$, $b = 4$, $K = 27$, distance-19 surface code with $n = 6840$), the spatial term dominates standard convolution ($\sim 77\%$ of MACs), so depthwise factorization---which reduces the spatial term from $nK(H/b)^2$ to $nK(H/b)$---yields a net $\sim 4\times$ speedup rather than the $\sim 64\times$ reduction in spatial cost alone, because the pointwise projections (unchanged) become the bottleneck.
Local attention costs roughly $1/3$ that of standard convolution, while full attention costs ${\sim}3.4\times$ more.
Convolution provides the strongest inductive bias (achieving the best accuracy in our ablation) at higher per-step cost, while local attention offers substantially reduced MACs with degraded accuracy at fixed model size.
A practical limitation of local attention is the lack of highly optimized CUDA kernels, making it slower to train and evaluate in practice despite its lower theoretical cost; as this ecosystem matures, local attention may become a compelling alternative for latency-constrained deployment.

\begin{figure*}
    \centering
    \includegraphics[width=0.9\textwidth]{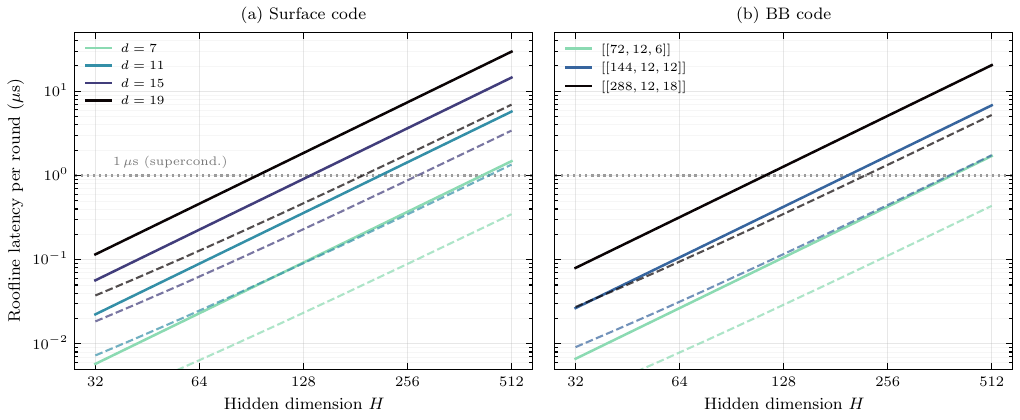}
    \caption{\textbf{Roofline latency per syndrome round on an AMD Versal AI Core FPGA (133 TOPS INT8).}
    Solid lines: standard convolution; dashed lines: depthwise convolution.
    Both panels assume network depth $L=d$ and bottleneck factor $b=4$.
    (a)~Surface codes with $K=27$ ($3\times 3\times 3$ kernel) and $n=d^2$ detectors per round.
    (b)~Bivariate bicycle codes with $K=24$ (two bipartite steps of 12 neighbors each) and $n$ equal to the number of physical qubits.
    The horizontal dotted line marks the $1\,\mu$s-per-round budget for superconducting qubits.
    Latency scales as $1/T$ for other hardware targets: a TPU v1 (92 TOPS) or Edge TPU (4 TOPS) would shift all curves upward by $1.4\times$ or $33\times$, respectively.}
    \refstepcounter{EDfig}\label{fig:roofline}
\end{figure*}

Extended Data Fig.~\ref{fig:roofline} shows the roofline latency per round as a function of hidden dimension $H$ for both surface codes and bivariate bicycle codes, assuming the peak throughput of an AMD Versal AI Core FPGA (133\,TOPS---tera MACs per second---in INT8 precision)~\cite{footnote_versal}.
FP8 and INT8 throughput are often comparable on modern tensor engines, so we use published INT8 throughput as a rough proxy for FP8 roofline estimates.
For reference, a Google TPU v1 provides 92 TOPS~\cite{jouppi2017datacenter} and a Google Edge TPU provides 4 TOPS at only $\sim 2$\,W; latency scales inversely with throughput.

Even the largest models are comfortably within the $\sim 1$\,ms decoding budget for trapped-ion and neutral-atom platforms.
For superconducting qubits ($\sim 1\,\mu$s budget), standard convolutions at $H=256$ require $\sim 10\,\mu$s per round---roughly one order of magnitude too slow.
However, depthwise convolutions at moderate widths ($H \leq 128$) approach the $1\,\mu$s target, and further reduction via lower precision (INT4) or smaller models could close the remaining gap.
BB codes have substantially lower latency than surface codes at the same hidden dimension, because both $n$ and $L$ are smaller: $n=144$ versus $n = d^2 = 361$ at $d=19$, and $L=12$ versus $L=19$.

We emphasize that these are \emph{roofline} estimates assuming 100\% hardware utilization.
On GPUs, single-sample (unbatched) inference is dominated by kernel launch overhead and memory transfers rather than compute, making GPU latency a poor proxy for dedicated hardware performance.
On FPGAs and ASICs, where the computation graph can be spatially mapped and weights stored on-chip, utilization approaching the roofline is achievable for regular, feed-forward architectures like ours.

\naturepar{Dataflow architecture}
\label{sec:dataflow}
The regularity of the convolutional architecture enables a spatial dataflow implementation in which the entire network is physically unrolled onto the FPGA or ASIC fabric.
Each layer is mapped to a dedicated hardware block, and data flows through the pipeline without dynamic scheduling or control flow---a direct consequence of the feed-forward, fixed-depth architecture.

Unlike the standard GPU execution model, where all $n$ detectors are processed through layer $l$ before layer $l+1$ begins, a spatially pipelined design maps each layer to a dedicated circuit block: as soon as a spatial tile exits layer $l$, layer $l+1$ begins processing it while layer $l$ handles the next tile.
Locality makes this possible---layer $l+1$ requires only a bounded neighborhood of layer $l$'s output (the \emph{halo}, one position wide for a $3\times 3\times 3$ kernel), not the entire activation tensor.
Within each stage, the residual connection requires buffering the block's input until the output is ready for the addition; at $n = 361$ positions and $H = 128$ in FP8, this is $\sim 46$\,KB per block, or $\sim 0.9$\,MB total for $L = 19$ layers.
With FP8 weights and batch normalization folded into the convolution parameters, weight storage is $K \times (H/b)^2 + 2H^2/b$ bytes per layer for a standard bottleneck convolution or $K \times H/b + 2H^2/b$ bytes for the depthwise variant: $\sim 0.7$\,MB or $\sim 0.2$\,MB total for a deployment-sized model ($H=128$, $K=27$, $b=4$, $L=19$).
Modern FPGAs provide 10--50\,MB of on-chip block RAM, so the entire model---weights, halos, and residual buffers---fits on-chip without external memory access, eliminating the memory bandwidth bottleneck that dominates GPU inference.

A further advantage over iterative decoders is deterministic latency.
Unlike belief propagation (which may require a variable number of iterations) or matching (which involves data-dependent graph traversal), the convolutional decoder has fixed-depth computation: every input traverses exactly the same computational path in the same number of clock cycles.
This is critical for real-time QEC systems, which are more sensitive to worst-case latency than average-case latency.

\end{document}